\documentclass[prl,10pt,showpacs,amsmath,twocolumn,floatfix]{revtex4}
\usepackage{amsmath}
\usepackage{amsfonts}
\usepackage{graphicx}

\newcommand{\diff}[2]{\frac{d #1}{d #2}}

\newcommand{\avg}[1]{\langle#1\rangle}

\newcommand{\bs}[1]{\boldsymbol{#1}}

\newcommand{\bk}[1]{\left(#1\right)}

\newcommand{\coup}{\kappa}
\newcommand{\decay}{\gamma}
\begin{document}

\title{Coherent Quantum-Noise Cancellation for Optomechanical Sensors}

\author{Mankei Tsang}

\email{mankei@unm.edu}

\affiliation{Center for Quantum Information and Control, University
of New Mexico, MSC07--4220, Albuquerque, New Mexico 87131-0001, USA}

\author{Carlton M.~Caves}

\affiliation{Center for Quantum Information and Control, University
of New Mexico, MSC07--4220, Albuquerque, New Mexico 87131-0001, USA}

\date{\today}

\begin{abstract}
  Using a flowchart representation of quantum optomechanical
  dynamics, we design coherent quantum-noise-cancellation schemes
  that can eliminate the back-action noise induced by radiation
  pressure at all frequencies and thus overcome the standard quantum
  limit of force sensing.  The proposed schemes can be regarded as
  novel examples of coherent feedforward quantum control.
\end{abstract}

\maketitle

Noise introduces error to communication or sensing because it is
unknown to the observer and cannot be distinguished from the desired
signal.  If the noise can be measured separately, one can physically
or computationally remove the noise.  This is the principle of noise
cancellation, a technique that has seen widespread commercial
application, especially for acoustic-noise control~\cite{hansen}:
noise-cancelling headphones, for example, work by recording the
ambient noise and playing it back with opposite amplitude to
interfere destructively with the noise reaching the inner ear.

Here we consider the application of noise cancellation to quantum
systems. We focus on optomechanical force sensors, in which quantum
radiation-pressure fluctuations on a moving mirror within an optical
cavity can introduce excess noise, called the back-action noise, in
addition to the shot noise at the cavity output.  This leads to the
standard quantum limit (SQL) on detecting a classical force on the
mirror~\cite{braginsky}.  Various methods of quantum-noise reduction
have been proposed to overcome the SQL, including frequency-dependent
squeezing of the input
light~\cite{unruh,bondurant84,kimble,unruh_exp,khalili10},
variational measurement~\cite{vyatchanin,kimble,khalili10},
introducing a Kerr medium inside the optical
cavity~\cite{bondurant86}, and the use of dual mechanical
resonators~\cite{briant,briant_exp} or an optical
spring~\cite{buonanno,buonanno_exp} to modify the mechanical response
function.  Several of these proposals have been
implemented~\cite{unruh_exp,briant_exp,buonanno_exp}, though not in
the quantum regime.  All can be regarded as
quantum-noise-cancellation (QNC) schemes, which utilize destructive
interference to reduce or eliminate the effects of back action. With
back action tamed, squeezing of the input light~\cite{caves} can be
used to improve sensitivity further.

To facilitate understanding of the QNC concept and design of new QNC
schemes, we introduce the use of flowcharts to depict the quantum
dynamics.
Using the flowcharts, we design a few novel QNC schemes that can
eliminate back-action noise at all frequencies and should require less
space than previous broadband QNC schemes.
Although the effects of quantum radiation-pressure noise are
only now becoming detectable~\cite{verlot09}, the field of
optomechanics has recently seen rapid progress~\cite{marquardt}.
Back-action noise is expected to become a major issue in future
optomechanical sensors. Combined with quantum filtering and
smoothing~\cite{tsang}, QNC has the potential to improve
significantly the performance of future force sensors beyond
conventional quantum limits. In the context of quantum
control~\cite{control}, QNC schemes can be regarded as examples
of \emph{coherent feedforward\/} quantum control, to be
contrasted with measurement-based feedforward control~\cite{lam}
and coherent feedback control~\cite{james} techniques.


\begin{figure}[htbp]
\centerline{\includegraphics[width=0.35\textwidth]{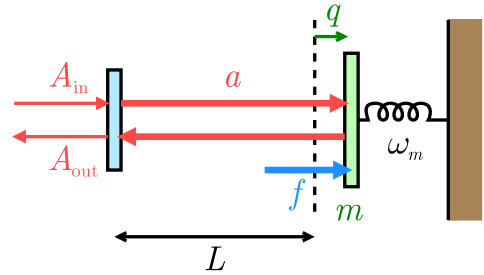}}
\caption{(Color online) Optomechanical force sensor.}
\label{setup}
\end{figure}

Consider now a Fabry-Perot cavity with a moving mirror, as depicted
in Fig.~\ref{setup}.  The mirror is modeled optically as perfectly
reflecting and mechanically as a harmonic oscillator with position
operator $q(t)$, momentum operator $p(t)$, mass $m$, and resonant
frequency $\omega_m$, subject to a force $f(t)$.  The cavity is
pumped with an input beam $A_{\rm in}(t)$, incident on the partially
transmitting mirror on the left.  We work in a rotating frame that
removes the harmonic time dependence at the input beam's carrier
frequency~$\omega_0$.  All phases are referenced to the input field,
which has constant (real) mean amplitude $\avg{A_{\rm in}}=\mathcal
A_{\rm in}$. The intra-cavity field decays at rate $\decay$ due to
coupling through the partially transmitting mirror to the output beam
$A_{\rm out}(t)$. The intra-cavity optical field $a(t)$ is assumed to
be resonant at the carrier frequency $\omega_0$ when the mirror is
displaced to its equilibrium position under the mean radiation
pressure, at which point the round-trip length is $2L$. The mirror
position $q(t)$ is defined relative to this equilibrium position. The
force is estimated by continuous homodyne measurement of an
appropriate quadrature of $A_{\rm out}$, labeled 2 in the following
and usually the phase quadrature.  The operators obey canonical
commutation relations: $[q(t),p(t)] = i\hbar$, $[a(t),a^\dagger(t)] =
1$, and $[A_{\rm in}(t),A_{\rm in}^\dagger(t')] = \delta(t-t')$.

One can generalize this basic setup to more elaborate configurations,
e.g., more complicated optical and mechanical mode structures or
detuned cavity excitation, which gives rise to an optical
spring~\cite{buonanno,buonanno_exp}.  A major difference from a
two-armed interferometer is that squeezed light must be input in the
same beam as the mean field that powers the system.  Nonetheless, this
minimal setup already captures the salient features of cavity
opto\-mechanical systems.

The operator equations of motion are
\begin{align}
\diff{q}{t} &= \frac{p}{m}\;,
\quad
\diff{p}{t} = -m\omega_m^2 q +  \frac{\hbar\omega_0}{L}
\bk{a^\dagger a-\alpha^2}+ f\;,
\nonumber\\
\diff{a}{t} &= -\decay a+\frac{i\omega_0}{L}qa+ \sqrt{2\decay}A_{\rm in}\;,
\quad
A_{\rm out} = \sqrt{2\decay}a-A_{\rm in}\;,
\label{A2}
\end{align}
where $\alpha=\mathcal A_{\rm in}\sqrt{2/\decay}$ is the mean cavity
field ($\avg{a^\dagger a}=\alpha^2$).   Removing mean fields and
defining amplitude and phase quadrature operators for the
fluctuations that remain, $A_{\rm in}=\mathcal A_{\rm
in}+(\xi_1+i\xi_2)/\sqrt2$, $A_{\rm out}=\mathcal A_{\rm
in}+(\eta_1+i\eta_2)/\sqrt2$, and $a=\alpha+(a_1+ia_2)/\sqrt2$, one
can linearize Eqs.~(\ref{A2}) by neglecting quadratic terms and
obtain a system of linear differential equations,
\begin{equation}
\diff{\bs x}{t}= \bs F \bs x + \bs G\bs w\;,
\qquad
\bs y= \bs H\bs x + \bs J \bs w\;.
\label{dx}
\end{equation}
Here the \emph{state variables\/} $\bs x \equiv
\begin{pmatrix}q&p&a_1&a_2\end{pmatrix}^T$, inputs $\bs w \equiv
\begin{pmatrix}f&\xi_1&\xi_2\end{pmatrix}^T$, and output signals
$\bs y \equiv\begin{pmatrix}\eta_1&\eta_2\end{pmatrix}^T$ are
related by the matrices
\begin{align}\label{FGHJ}
\bs F&\equiv
\begin{pmatrix}
0 & 1/m & 0 & 0\\
-m\omega_m^2 & 0 & \hbar \coup & 0\\
0 & 0 & -\decay & 0\\
\coup & 0 & 0 & -\decay
\end{pmatrix}\;,
\quad
\bs J\equiv
\begin{pmatrix}
    0 & -1 & 0\\
    0 & 0 & -1
\end{pmatrix}\;,
\nonumber\\
\bs G&\equiv
\begin{pmatrix}
0 & 0 & 0\\
1 & 0 & 0\\
0 & \sqrt{2\decay} & 0\\
0 & 0 & \sqrt{2\decay}
\end{pmatrix}\;,
\quad
\bs H\equiv
\begin{pmatrix}
0 & 0 & \sqrt{2\decay} & 0\\
0 & 0 & 0 & \sqrt{2\decay}
\end{pmatrix}\;,
\end{align}
where $\coup\equiv\sqrt2\alpha\omega_0/L$ is an optomechanical
coupling strength.

\begin{figure}[htbp]
  \centerline{\includegraphics[width=0.45\textwidth]{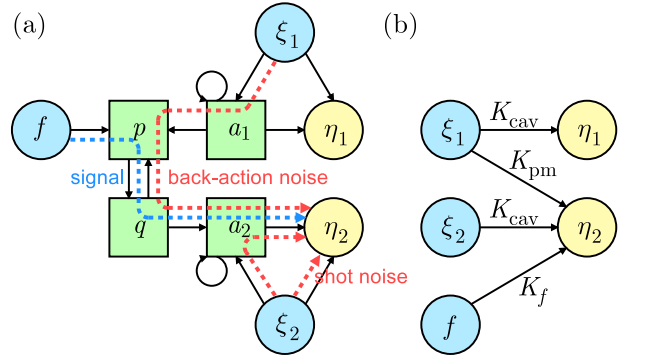}}
  \caption{(Color online) (a)~Flowchart representation of
    Eqs.~(\ref{dx}) and~(\ref{FGHJ}).  (b)~Simplified flowchart
    depicting only the input and output signals linked by
    transfer functions.}
\label{flowchart}
\end{figure}

In control-system design, it is often illuminating to draw a
block diagram to represent a system of differential equations
\cite{franklin}.  Here we use a simpler depiction, which we call
a flowchart, as shown in Fig.~\ref{flowchart}(a) for
Eqs.~(\ref{dx}) and (\ref{FGHJ}). Arrows point from a variable
on the right-hand side of an equation of motion to a connected
variable on the left-hand side.  Since the system is linear, the
output signals are sums of independent contributions from the
inputs, so one can easily depict the flow of signal and noise
from input to output.  In Fig.~\ref{flowchart}(a), the dashed
arrows depict contributions to the output quadrature~$\eta_2$
from the signal $f$ and from the input amplitude and phase
fluctuations, $\xi_1$ and $\xi_2$.  The $\xi_2$ contribution is
commonly known as shot noise, while the $\xi_1$ contribution is
the back-action noise due to the Kerr-like ponderomotive
coupling of the cavity amplitude quadrature $a_1$ to the phase
quadrature~$a_2$.


The solution for the state variables contains a transient solution,
which decays exponentially and which can be ignored by pushing the
initial time to $-\infty$.  It is convenient to Fourier transform the
remaining inhomogenous solution, writing $\bs y(\Omega) = \bs
K(\Omega)\bs w(\Omega)$, where the overall transfer matrix is
\begin{equation}
\bs K(\Omega)=-\bs H(i\Omega\bs I+\bs F)^{-1}\bs G + \bs J
=\begin{pmatrix}0 & K_{\rm cav} & 0\\
K_f & K_{\textrm{pm}} & K_{\rm cav}
\end{pmatrix}\;,
\end{equation}
with $\bs I$ being the identity matrix.  The cavity and ponderomotive
transfer functions are
\begin{align}
K_{\rm cav}(\Omega) &=
\frac{\decay+i\Omega}{\decay-i\Omega}\;,
&
K_{\rm pm}(\Omega) &=
\frac{2\decay\hbar \coup^2/m}{(\omega_m^2-\Omega^2)(\decay-i\Omega)^2}\;,
\end{align}
and the signal transfer function is $K_f(\Omega)= K_{\rm
pm}(\Omega)(\decay-i\Omega)/\hbar\coup\sqrt{2\decay}$.  A
simplified flowchart that connects outputs to inputs by transfer
functions is shown in Fig.~\ref{flowchart}(b).

\begin{figure}[htbp]
\centerline{\includegraphics[width=0.48\textwidth]{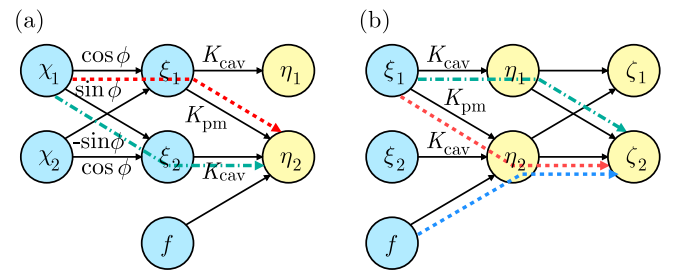}}
\caption{(Color online) (a)~Frequency-dependent input
squeezing. (b)~Variational-measurement scheme.}
\label{unruh}
\end{figure}

Unruh's proposal for frequency-dependent input
squeezing~\cite{unruh,bondurant84,kimble,unruh_exp,khalili10}
and Vyatchanin and Matsko's variational-measurement scheme
\cite{vyatchanin,kimble,khalili10} can be understood using
flowcharts. First consider Unruh's proposal, flowcharted in
Fig.~\ref{unruh}(a). The input beam, with quadratures $\chi_1$
and $\chi_2$, is processed through a passive system that rotates
the quadratures by a frequency-dependent angle $\phi(\Omega)$,
after which the beam is displaced by $\mathcal A_{\rm in}$ to
produce the input field $A_{\rm in}$ to the optical cavity.  The
transfer functions from $\chi_1$ and $\chi_2$ to the output
phase quadrature $\eta_2$ are
$K_{\eta_2\chi_1}=K_{\textrm{pm}}\cos\phi+
K_{\textrm{cav}}\sin\phi$ and
$K_{\eta_2\chi_2}=-K_{\textrm{pm}}\sin\phi+K_{\textrm{cav}}\cos\phi$.
The rotation is chosen to make the contribution from $\chi_1$
vanish, i.e., $K_{\eta_2\chi_1}=0$, by introducing destructively
interfering paths from $\chi_1$ to $\eta_2$, as shown in
Fig.~\ref{unruh}(a). This requires $\tan\phi(\Omega)=-K_{\rm
pm}(\Omega)/K_{\rm cav}(\Omega)$.
Kimble~\textit{et al.}~\cite{kimble} devised a method for performing
the dispersive rotation over a wide bandwidth by filtering the input
beam through two detuned cavities.  If the input to the filtering
cavities is vacuum, the rotation has no effect on sensitivity; since
$\chi_2$ is now entirely responsible for the output noise, however, it
can be squeezed to improve the sensitivity to the shot-noise limit and
beyond.

The variational-measurement scheme can be understood using the
flowchart in Fig.~\ref{unruh}(b).  This scheme dispersively rotates
the output light and measures the final output quadrature $\zeta_2$,
which is a combination of the original output quadratures $\eta_1$ and
$\eta_2$.  This introduces an ``anti-noise'' path from $\xi_1$ to
$\zeta_2$ via $\eta_1$, which can be used to cancel the original
back-action noise path via $\eta_2$.  Given the similarity of the
flowcharts in Fig.~\ref{unruh}, the required dispersive rotation is
the same as that for Unruh's proposal and can again be implemented
using two detuned cavities.  With back-action noise eliminated, the
sensitivity is limited by shot noise and can be improved further by
squeezing the input quadrature~$\xi_2$~\cite{kimble}.

\begin{figure}[htbp]
\centerline{\includegraphics[width=0.45\textwidth]{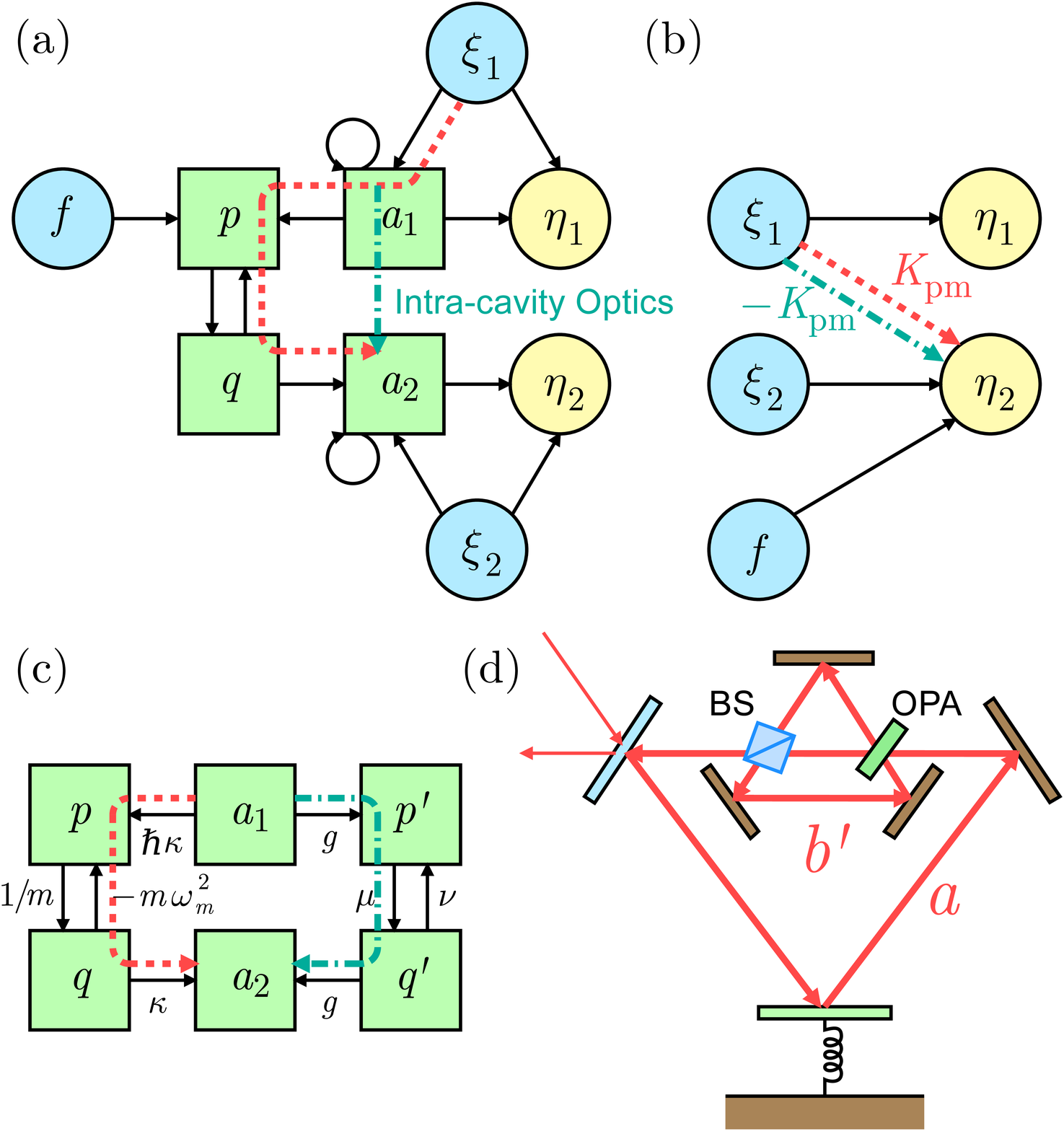}}
\caption{(Color online) (a)~Back-action noise cancellation achieved
  by introducing an anti-noise path (green dash-dot arrow)
  from the intra-cavity amplitude quadrature $a_1$ to the phase
  quadrature $a_2$. (b)~Simplified flowchart.  (c)~Detailed flowchart
  of the ponderomotive coupling and the intra-cavity matched squeezing.
  (d)~An implementation of the matched squeezing scheme.}
\label{cancel}
\end{figure}

While the aforementioned schemes can be regarded as examples of
coherent QNC, it is clear from Fig.~\ref{flowchart} that a more
direct way of cancelling the back-action noise is to introduce
an anti-noise path from $a_1$ to $a_2$, as illustrated in
Fig.~\ref{cancel}.  This calls for coherent processing of the
intra-cavity field, using parametric interactions to undo the
ponderomotive squeezing.  The use of a Kerr
medium~\cite{bondurant86}, the dual-mechanical-resonator
scheme~\cite{briant,briant_exp}, and even an optical
spring~\cite{buonanno,buonanno_exp} can also be thought of as
intra-cavity QNC schemes, but they cannot eliminate the
back-action noise at all frequencies.  To achieve broadband QNC,
we introduce two \emph{auxiliary\/} state variables, $q'$ and
$p'$, which play the role of $q$ and $p$ in the anti-noise path,
as shown in Fig.~\ref{cancel}(c).  Adjustable parameters $g$,
$\mu$, and $\nu$, all with units of frequency, characterize the
couplings in the anti-noise path. The anti-noise transfer
function, indicated in Fig.~\ref{cancel}(b) by the green
dash-dot arrow, is $2\decay\mu
g^2/(-\mu\nu-\Omega^2)(\decay-i\Omega)^2=-K_{\textrm{pm}}$.  We
need $\mu\nu=-\omega_m^2$ and $\mu g^2=-\hbar \coup^2/m$, thus
requiring $\mu$ to be negative.  This cannot be implemented by
ponderomotive coupling to another mechanical oscillator unless
its mass is negative.

The matched squeezing can nonetheless be achieved by coupling the
intra-cavity field to an auxiliary field $b\equiv(q'+ip')/\sqrt2$.
The required parameters are $\nu=-\mu=\omega_m$ and
$g=\coup\sqrt{\hbar/m\omega_m}=\alpha\omega_0\sqrt{2\hbar/m\omega_m}/L$,
and the equations for $a$ and $b$ become
\begin{align}
\diff{a}{t} &= -\decay a+\frac{i\omega_0}{L}q a
+ \frac{ig}{2}(b+b^\dagger) + \sqrt{2\decay}A_{\rm in}\;,
\nonumber\\
\diff{b}{t} &=
i\omega_m b + \frac{ig}{2}(a + a^\dagger)-ig\alpha\;.
\label{motion2}
\end{align}
The auxiliary field should be inside another optical cavity,
with resonant frequency $\omega_0-\omega_m$; it plays the role
of a \emph{negative-energy\/} mode in the anti-noise path.  The
constant driving term $-ig\alpha$, which removes the mean field
from $b$, can be eliminated by redefining the auxiliary mode as
$b'=b-g\alpha/\omega_m$ and displacing the input field
$A_{\textrm{in}}$, resulting in the following equations of
motion:
\begin{align}
\diff{a}{t} &= -\decay a+\frac{i\omega_0}{L}q a
+ \frac{ig}{2}(b'+b^{\prime\dagger}) + \sqrt{2\decay}A_{\rm in}'\;,
\nonumber\\
\diff{b'}{t} &=
i\omega_m b' + \frac{ig}{2}(a + a^\dagger)\;,
\quad
A'_{\rm in} = A_{\rm in}+i\mathcal{A}_{\rm in}\frac{g^2}{\decay\omega_m}\;.
\label{motion3}
\end{align}
The coupling between $a$ and $b'$ can then be realized using a
beamsplitter (BS) and an optical parametric amplifier (OPA), as
schematically shown in Fig.~\ref{cancel}(d).  With the
back-action noise removed, only the shot noise from $\xi_2$
remains, and one can improve sensitivity further by increasing
the optical power or by squeezing $\xi_2$.


In terms of the coupling constant $g$, the ponderomotive
transfer function has the form
\begin{equation}
K_{\rm pm}(\Omega)=
\frac{g^2}{\decay\omega_m}
\frac{\omega_m^2}{\omega_m^2-\Omega^2}
\frac{2\decay^2}{(\decay-i\Omega)^2}\;,
\end{equation}
so $|K_{\rm pm}(\Omega)|/|K_{\rm cav}(\Omega)|\sim g^2/\decay\omega_m$
for frequencies away from the mechanical resonance; back action is
thus important when $g\agt\sqrt{\decay\omega_m}$.  The required
single-pass idler gain of the OPA is thus $G\simeq (gL/c)^2 \agt
\decay\omega_m (L/c)^2\equiv G_0$. $G_0$ ranges from about
$(2\pi\times 10~\textrm{Hz}\times5~\textrm{km}/c)^2\simeq 10^{-6}$ for
gravitational-wave detectors to $(2\pi\times
50~\textrm{MHz}\times100~\mu\textrm{m}/c)^2\simeq10^{-8}$ for
microscale systems~\cite{marquardt}, so the required $G$ should be
easily achievable with current OPA technology.

\begin{figure}[htbp]
  \centerline{\includegraphics[width=0.48\textwidth]{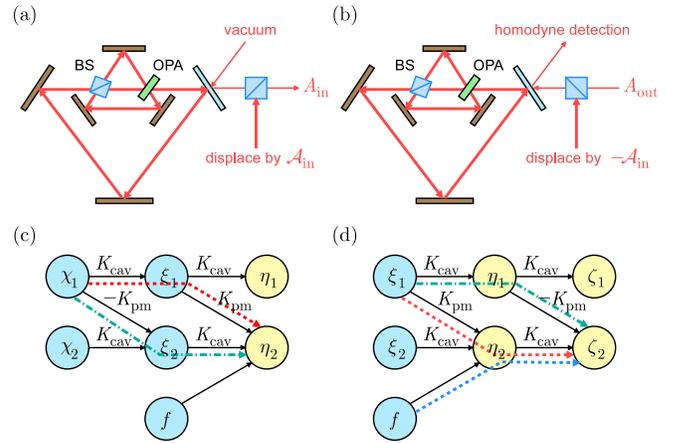}}
  \caption{(Color online) Implementations of the (a)~input and
    (b)~output matched squeezing schemes and the associated flowcharts~(c)
    and~(d). For a two-armed interferometer, one only needs to
    squeeze the dark input or output port and does not need to perform
    the displacement.}
  \label{input_output_match}
\end{figure}

There is another potential problem: the number of photons in the
auxiliary cavity in Fig.~\ref{cancel}(d) is $\avg{b'^\dagger b'}
= g^2\alpha^2/\omega_m^2$, which becomes significantly higher
than that in the primary cavity when $\omega_m$ is small, making
the intra-cavity scheme problematic for applications that use
high circulating power.  This problem can be alleviated by using
matched squeezing to modify the input or output optics.  This
requires the use of another double-cavity setup, similar to that
in Fig.~\ref{cancel}(d), but without the moving mirror, to
pre-squeeze the input light going into the sensor cavity or
post-squeeze the output light, as shown in
Figs.~\ref{input_output_match}(a) and~(b). The flowcharts for
these schemes, shown in Figs.~\ref{input_output_match}(c)
and~(d), demonstrate broadband cancellation of back-action noise
much like the intra-cavity scheme. If one is interested only in
low frequencies $\Omega \ll\gamma$, the input/output matched
squeezing can be implemented using smaller cavities with a
larger decay rate $\gamma$ and a larger coupling constant $g$,
with $g^2/\decay$ held constant.

We have assumed an ideal system, neglecting intrinsic mechanical
and optical losses, to illustrate the essential features of
\hbox{QNC}. Such assumptions are unrealistic in practice.  To
assess QNC schemes, it will be important to take into account
the fluctuations associated with dissipation, such as the
thermal noise associated with mechanical damping.  The design
and performance of QNC in the presence of realistic dissipation
and noise deserve further investigation.

We acknowledge discussions with T.~Kippenberg and S.~Waldman.
This work was supported in part by NSF Grant Nos.~PHY-0903953
and~PHY-0653596 and ONR Grant No.~N00014-07-1-0304.



\begin{thebibliography}{}

\bibitem{hansen}C.~H.\ Hansen,
\textit{Understanding Active Noise Cancellation\/} (Taylor \&
Francis, London, 2001).



\bibitem{braginsky}V.~B.\ Braginsky and F.~Ya.\ Khalili,
\textit{Quantum Measurement\/} (Cambridge University Press,
Cambridge, 1992).

\bibitem{unruh}W.~G.\ Unruh,
in \textit{Quantum Optics, Experimental Gravitation, and Measurement Theory},
edited by P.\ Meystre and M.\ O.\ Scully (Plenum, New York, 1982), p.~647.

\bibitem{bondurant84}R.~S. Bondurant and J.~H. Shapiro, \prd \textbf{30},
2548 (1984); M.~T. Jaekel and S.~Reynaud, Europhys. Lett. {\bf 13},
301 (1990); A.~Luis and L.~L. S{\'a}nchez-Soto, \pra \textbf{45},
8228 (1992).

\bibitem{kimble}H.~J.\ Kimble \textit{et al.},
\prd \textbf{65}, 022002 (2001).

\bibitem{unruh_exp}C.\ M.\ Mow-Lowry~\textit{et al.},
\prl \textbf{92}, 161102 (2004).

\bibitem{khalili10}F.~Ya.~Khalili, e-print {\tt arXiv:1003.2859}, and
  references therein.

\bibitem{vyatchanin}S.~P. Vyatchanin and A.~B. Matsko, JETP
\textbf{77}, 218 (1996); S.\ P.\ Vyatchanin and E.\ A.\ Zubova,
Phys.\ Lett.\ A \textbf{201}, 269 (1995).

\bibitem{bondurant86}R.\ S.\ Bondurant, \pra \textbf{34}, 3927 (1986).

\bibitem{briant}T.~Briant~\textit{et al.}, \prd \textbf{67}, 102005 (2003).

\bibitem{briant_exp}T.\ Caniard~\textit{et al.}, \prl \textbf{99}, 110801 (2007).

\bibitem{buonanno}A.~Buonanno and Y.~Chen, \prd \textbf{64}, 042006 (2001);
\textbf{65}, 042001 (2002).

\bibitem{buonanno_exp}P.~Verlot~\textit{et al.}, \prl \textbf{104}, 133602 (2010).

\bibitem{caves}C.~M.\ Caves, \prd \textbf{23}, 1693 (1981).

\bibitem{verlot09}P.\ Verlot \textit{et al.}, \prl \textbf{102}, 103601
(2009).

\bibitem{marquardt}T.~J.\ Kippenberg and K.~J.\ Vahala, Science \textbf{321},
1172 (2008); F.~Marquardt and S.~M.\ Girvin, Physics \textbf{2}, 40
(2009).

\bibitem{tsang}M.~Tsang, \prl \textbf{102}, 250403 (2009).



\bibitem{control}
H.~Mabuchi and N.~Khaneja, Int.\ J.\ Robust Nonlinear Control
\textbf{15}, 647 (2005), and references therein.

\bibitem{lam}
P.~K.\ Lam \textit{et al.}, \prl \textbf{79}, 1471 (1997);
U.\ L.\ Andersen and R.~Filip, in \textit{Progress in Optics},
Vol.~53, edited by E.~Wolf (Elsevier, Amsterdam, 2009), p.~365, and
references therein.

\bibitem{james}M.\ R.\ James, H.\ I.\ Nurdin, and I.\ R.\ Petersen,
IEEE Trans.\ Auto.\ Control, \textbf{53}, 1787 (2008).

\bibitem{franklin}G.~F.\ Franklin, J.~D.\ Powell, and A.~Emami-Naeini,
\textit{Feedback Control of Dynamic Systems} (Prentice Hall, Upper
Saddle River, 2002).







\end{thebibliography}
\end{document}